\documentclass[journal]{IEEEtran}

\usepackage{placeins}
\usepackage{amsmath}
\usepackage{algorithm}
\usepackage{algorithmic}
\usepackage{booktabs}
\usepackage{multirow}
\usepackage[utf8]{inputenc} 
\usepackage[T1]{fontenc}    
\usepackage{hyperref}       
\usepackage{url}            
\usepackage{booktabs}       
\usepackage{amsfonts}       
\usepackage{nicefrac}       
\usepackage{microtype}      
\usepackage{xcolor}         
\usepackage{graphicx}
\usepackage{float}
\usepackage[numbers,sort&compress]{natbib}
\usepackage{caption} 

\newfloat{figtab}{htb}{fgtb}
\makeatletter
  \newcommand\figcaption{\def\@captype{figure}\caption}
  \newcommand\tabcaption{\def\@captype{table}\caption}
\makeatother

\def\BibTeX{{\rm B\kern-.05em{\sc i\kern-.025em b}\kern-.08em
    T\kern-.1667em\lower.7ex\hbox{E}\kern-.125emX}}

\title{iFuzzyTL: Interpretable Fuzzy Transfer Learning for SSVEP BCI System}
\label{sec:title}
\author{
\IEEEauthorblockN{Xiaowei Jiang\textsuperscript{1}\textsuperscript{\dag}, Beining Cao\textsuperscript{1}\textsuperscript{\dag}, Liang Ou\textsuperscript{1}, Yu-Cheng Chang\textsuperscript{1}, Thomas Do\textsuperscript{1}, Chin-Teng Lin\textsuperscript{*}\textsuperscript{1}\\}

\IEEEauthorblockA{\textsuperscript{1}GrapheneX-UTS Human-centric AI Centre, Australian AI Institute, School of Computer Science,\\ Faculty of Engineering and Information Technology, University of Technology Sydney}
\thanks{\textsuperscript{\dag}Xiaowei Jiang and Beining Cao contributed equally to this work.}
\thanks{\textsuperscript{*}Corresponding author: Chin-Teng Lin. Email: chin-teng.lin@uts.edu.au}
}

\begin{document}
\maketitle

\begin{abstract}
The rapid evolution of Brain-Computer Interfaces (BCIs) has significantly influenced the domain of human-computer interaction, with Steady-State Visual Evoked Potentials (SSVEP) emerging as a notably robust paradigm. This study explores advanced classification techniques leveraging interpretable fuzzy transfer learning (iFuzzyTL) to enhance the adaptability and performance of SSVEP-based systems. Recent efforts have strengthened to reduce calibration requirements through innovative transfer learning approaches, which refine cross-subject generalizability and minimize calibration through strategic application of domain adaptation and few-shot learning strategies. Pioneering developments in deep learning also offer promising enhancements, facilitating robust domain adaptation and significantly improving system responsiveness and accuracy in SSVEP classification. However, these methods often require complex tuning and extensive data, limiting immediate applicability. iFuzzyTL introduces an adaptive framework that combines fuzzy logic principles with neural network architectures, focusing on efficient knowledge transfer and domain adaptation. iFuzzyTL refines input signal processing and classification in a human-interpretable format by integrating fuzzy inference systems and attention mechanisms. This approach bolsters the model's precision and aligns with real-world operational demands by effectively managing the inherent variability and uncertainty of EEG data. The model's efficacy is demonstrated across three datasets: 12JFPM (89.70\% accuracy for 1s with an information
transfer rate (ITR) of 149.58), Benchmark (85.81\% accuracy for 1s with an ITR of 213.99), and eldBETA (76.50\% accuracy for 1s with an ITR of 94.63), achieving state-of-the-art results and setting new benchmarks for SSVEP BCI performance.
\end{abstract}


\section{Introduction}

\IEEEPARstart{B}{rain-computer} interfaces (BCIs) have become increasingly popular in human-computer interaction (HCI) due to their intuitive nature\cite{guo2022ssvep, chen2020combination, cao2022building, lin2020direct}. BCIs allow for direct extraction of user intentions from the brain, bypassing the peripheral nervous system and muscle tissue\cite{samejima2021brain}. Among the various non-invasive EEG-based BCI paradigms, such as steady-state visual evoked potentials (SSVEP)\cite{cheng2002design}, P300\cite{cecotti2010convolutional}, and motor imagery (MI)\cite{schlogl2005characterization}, SSVEP is particularly noted for its high accuracy and robustness. In SSVEP BCIs, users focus on visual stimuli flickering at different frequencies, and their intent is deciphered by identifying the frequency of the observed flicker. Remarkably, research in this field has advanced to where forty commands can be distinguished within just one second of EEG data\cite{liu2020beta}.

\begin{figure}[htp]
    \centering
    \includegraphics[width=0.8\linewidth]{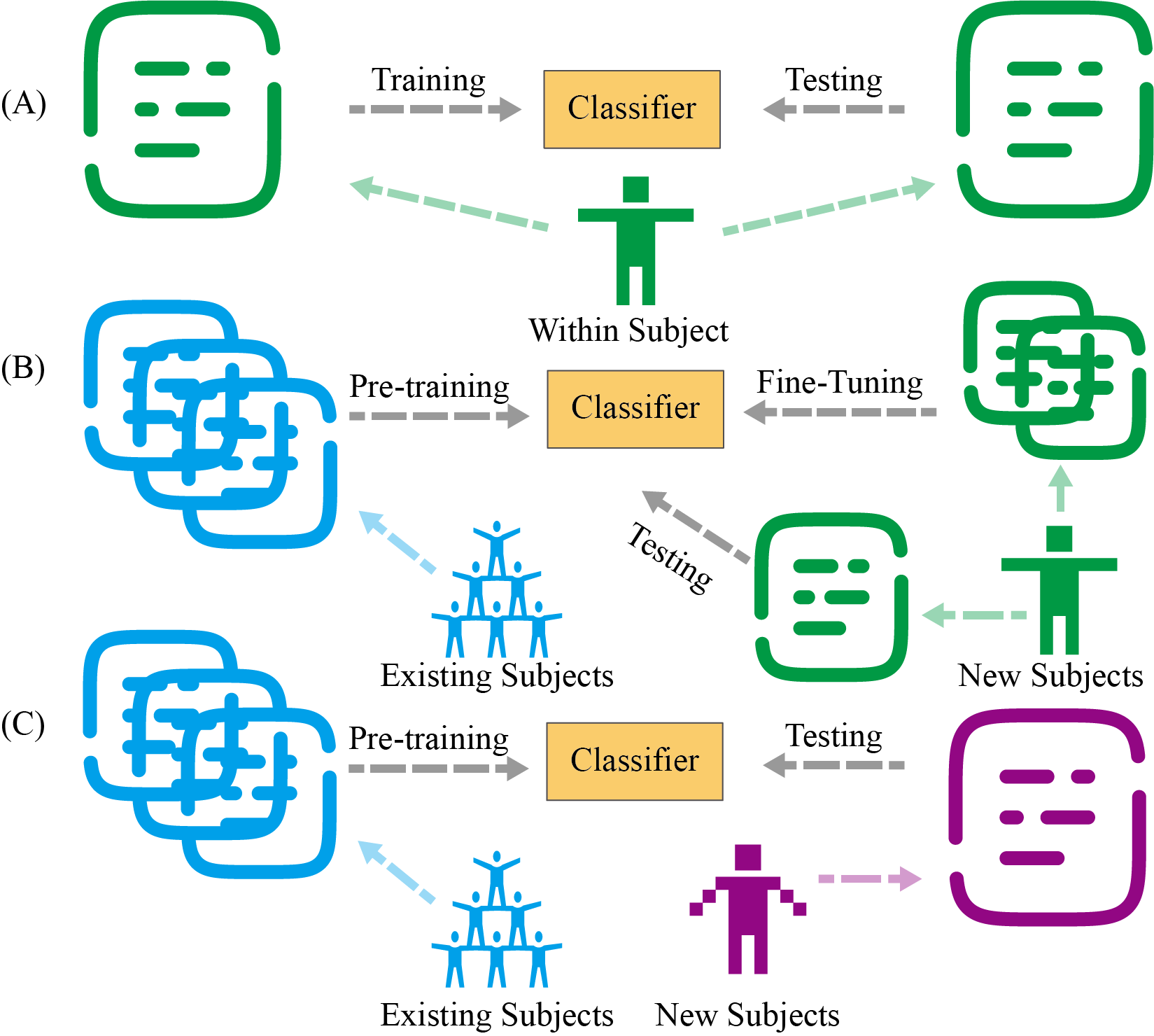}
    \caption{The diagram of three classification scenarios. 
    \textbf{(A)} intra-subject classification; 
    \textbf{(B)} inter-subject few-shot classification;
    \textbf{(C)} inter-subject zero-shot classification.
    }
    \label{fig:TLdemo}
\end{figure}

Classification methods for SSVEP are broadly categorized into unsupervised and supervised techniques. Canonical correlation analysis (CCA)\cite{lin2007frequency} and Filter-bank CCA\cite{chen2015filter} are traditional unsupervised methods that determine the target frequency by measuring the correlation between EEG signals and predefined reference signals. Although effective, their performance lags behind supervised methods, particularly with shorter EEG segments\cite{yuan2015enhancing} in the intra-subject classification task, as shown in Figure \ref{fig:TLdemo}(A). Consequently, supervised methods such as extended CCA (eCCA)\cite{nakanishi2014high}, task-related component analysis (TRCA)\cite{nakanishi2017enhancing}, and complex-spectrum convolutional neural networks (CCNN)\cite{ravi2020comparing} have been developed, significantly outperforming unsupervised approaches. However, these high-performing supervised methods require extensive data collection for model training or user-specific calibration, hindering their immediate usability\cite{luo2023almost, mai2023calibration}. As a result, transfer learning has emerged as a key research area in SSVEP studies\cite{chiang2021boosting, bassi2021transfer}.

This approach leverages knowledge gained from the source subjects to improve performance on new subjects \cite{panSurveyTransferLearning2010}, minimizing the need for extensive calibration typically required for personalized BCIs. Techniques such as domain adaptation are employed to modify models developed on one individual's data for use with another's, significantly enhancing cross-subject generalizability\cite{jayaramTransferLearningBrainComputer2016, wangReviewTransferLearning2015}. In SSVEP, some transfer learning methods based on CCA using the spatial filter and templates to learn the knowledge from existing domains \cite{chiangBoostingTemplatebasedSSVEP2021, yuan2015enhancing} Additionally, fine-tuning deep learning models pre-trained on large datasets in a domain-specific manner can substantially reduce the discrepancy between training and implementation environments, offering a robust solution for SSVEP BCI applications\cite{chen2023transformer, ravi2020comparing}. Hybrid strategies that combine classical signal processing with advanced machine learning techniques also play a crucial role. These methods preprocess EEG signals to extract features more invariant across subjects before training classifiers, thereby balancing performance with computational efficiency essential for real-time applications.

Recent advances have focused on reducing the need for calibration through few-shot learning approaches, as shown in Figure \ref{fig:TLdemo}(B). Pioneering work by Chi Man Wong et al. introduced a subject transfer-based CCA (stCCA), which utilizes cross-subject spatial filters and SSVEP templates to enhance transferability\cite{wong2020inter}. This method achieved an impressive information transfer rate of 198.18 $\pm$ 59.12 bits/min with minimal calibration trials for a 40-target task, Benchmark\cite{liu2020beta}. Further, numerous modified CCA methods have been proposed to refine few-shot learning in SSVEP\cite{lan2023cross, wang2021cross, wong2021online, du2024enhancing}. Alternatively, deep learning (DL) frameworks are renowned for their efficacy in utilizing previously acquired knowledge to address challenges in transfer learning and domain adaptation, effectively managing uncertainties and enhancing predictive accuracy across related domains \cite{panSurveyTransferLearning2010, luTransferLearningUsing2015}. In SSVEP, DL is also being explored for their potential in few-shot SSVEP transfer learning, such as convolutional neural network (CNN)\cite{xiong2024deep, liu2024convolutional} or Transformer-based\cite{chen2023transformer}, although they still necessitate some degree of calibration.

To entirely eliminate the reliance on calibration, novel zero-shot learning methods for SSVEP have been developed, as shown in Figure \ref{fig:TLdemo}(C). These include signal correlation analysis (CA) methods such as CCA and TRCA, which leverage parameters solely from the source domain\cite{huang2023cross, wang2023improving, bassi2021transfer}. While zero-shot CA methods generally underperform compared to their few-shot counterparts, DL-based transfer learning techniques, using architectures like long short-term memory (LSTM) and Transformers, have shown promising results in achieving higher accuracy\cite{guney2023transfer, liu2024convolutional}. Despite these advancements, DL methods are often criticized for their lack of interpretability compared to the transparent calculations of CA methods\cite{vinuesa2021interpretable}. Interpretability helps explain model failures and enhances system stability, therefore, developing an interpretable DL framework that elucidates the underlying mechanisms remains a critical challenge in the field.

The Fuzzy Neural Network (FNN) \cite{lin1996neural, 10183374,106218} stands out as a potential framework that combines the robustness of neural networks with the clarity of fuzzy logic systems. FNNs utilize fuzzy rules and membership functions to process inputs, thereby maintaining a logical structure that is both transparent and intuitive. This method contrasts sharply with more opaque models, clearly visualizing how inputs are transformed into outputs through human-understandable rules, thus helping user optimize the BCI system. Drawing on the principles of fuzzy logic \cite{luFuzzyMachineLearning2024}, particularly the Takagi–Sugeno–Kang (TSK) inference systems \cite{shihabudheen2018recent}, our work introduces the interpretable fuzzy transfer learning (iFuzzyTL) model—a novel Fuzzy Inference Systems (FISs) based on fuzzy set theory \cite{shihabudheen2018recent}, tailored for the SSVEP task. FISs have been further developed into a neural network architecture known as FNN, which can be trained using gradient descent optimization. The advantage of this architecture is that the fuzzy logic follows human intuitive deduction, providing good interpretability. A recent study, KAN \cite{liu2024kan}, indicates that by learning the activation function specific to each dimension, the network gains "internal degrees of freedom." This concept is embodied in the TSK model through the parameters $m_{d,r}$ and $\sigma_{d,r}$, which tailor unique bell-shaped activation functions for each $x_d$. This approach starkly contrasts linear-based models such as Transformers and CNNs, where each $x_d$ is simply multiplied by a corresponding weight. One paper demonstrates that introducing a Fuzzy Attention Layer significantly enhances the network's approximation capabilities by leveraging these internal degrees of freedom \cite{jiang2024fuzzybasedapproachpredicthuman}. 
Our model is crafted to exploit extensive experimental evidence from prior studies, serving as a robust knowledge base that enables significant adaptation to environments characterized by limited data availability \cite{zuoFuzzyRegressionTransfer2017}. Furthermore, fuzzy rule-based transfer learning models, including ours, have demonstrated remarkable capabilities in addressing the challenges posed by small source datasets in transfer learning scenarios, ensuring reliable performance even when existing data resources are sparse \cite{zuoFuzzyRuleBasedDomain2019, liuUnsupervisedHeterogeneousDomain2018, luFuzzyMultipleSourceTransfer2020, zhuDesignDevelopmentGranular2023, zuoFuzzyTransferLearning2019}, especially the application in brain signal processing \cite{chang2021exploring,reddy2021joint,hu2021fuzzy,zarandi2011systematic} and EEG-based BCI system \cite{chenEEGbasedTSKFuzzy2024}.

In addressing the challenges of domain adaptation, iFuzzyTL modifies the source domain model's input and/or output spaces through spatial transformations. This ensures that the fuzzy rules align more precisely with the target data, enhancing the model's robustness even with minimal available data. Furthermore, the capacity of fuzzy logic to cluster data and facilitate the separation of classes during the domain transfer process has been proved by unsupervised transfer learning models \cite{liuUnsupervisedHeterogeneousDomain2018, zuoGranularFuzzyRegression2018, tahmoresnezhadVisualDomainAdaptation2017}. Following the idea of clustering, iFuzzyTL calculates the membership degree based on the distance between input features and the centroid of fuzzy sets. Each centroid represents a prototype characteristic of its cluster, and the distances are measured using a suitable metric, typically Euclidean\cite{1980317}. The closer an input feature is to a fuzzy centroid, the higher its membership grade is to that centroid. The membership grade determines the firing strength of the fuzzy rules associated with the corresponding center, with the rule strength computed using a fuzzy operation (e.g., sum, product, min, or max) applied to the input membership grades. This approach enables the system to process inputs that exhibit varying degrees of similarity to known categories, and the nonlinearity provided by the Gaussian membership functions makes the approximation of real-world data more robust \cite{zhang2023robust}, thereby accommodating real-world data's inherent uncertainty and fuzziness. As a result, iFuzzyTL provides a robust, human-interpretable, and adaptable framework for applications requiring nuanced decision-making processes.

To further refine the model's capabilities, the dual-filter structure, which includes both spatial and temporal filters as applied by EEGNET \cite{lawhernEEGNetCompactConvolutional2018}, demonstrates significant enhancements in processing EEG data. iFuzzyTL incorporates Fuzzy Attention Layers \cite{jiang2024fuzzybasedapproachpredicthuman} as spatial and temporal filters to capture and generalize the central fuzzy rules within the network. This architecture effectively learns the domain knowledge of both spatial and temporal dependencies in the brain signals, enabling more accurate and robust feature extraction and domain adaptation, especially in transfer learning scenarios. These filters in iFuzzyTL integrate fuzzy set theory with neural network architectures to model SSVEP signal sequences as fuzzy sets, producing firing strengths in an \(S \times N\) matrix format. This approach parallels the mechanism of vanilla dot-product self-attention \cite{NIPS2017_3f5ee243, cheng2016long, paulus2018a}, enhancing the robustness and flexibility of the model in neurophysiological applications. By melding fuzzy logic with advanced attention mechanisms, iFuzzyTL facilitates efficient knowledge transfer across varying domains and sets a new benchmark in the field of computational intelligence-based transfer learning, especially for tasks involving complex signal patterns like SSVEP. Our model achieves the highest ITR and accuracy in three datasets as zero-shot learning, 12JFPM(89.70\% for 1s with ITR=149.58), Benchmark(85.81\% for 1s with ITR=213.99), and eldBETA(76.50\% for 1s with ITR=94.63), and is the State Of The Art (SOTA) model in the SSVEP transfer learning issue. We also demonstrate how the iFuzzyTL model enhances interpretability by revealing the temporal dynamics of firing strength and its harmonic relationships with target frequencies in the SSVEP task.

The contributions of this study are outlined as follows:
\begin{enumerate}
    \item \textbf{Development of iFuzzyTL:} We propose a fuzzy logic-based attention mechanism, named iFuzzyTL, which improves transferability in SSVEP tasks. This innovation minimizes the need for user-specific calibration and facilitates a plug-and-play experience for BCI systems.
    
    \item \textbf{Interpretable Framework Enhancement:} We introduce an interpretable approach that utilizes a human-understandable center for clustering and learning knowledge from the source domain. By integrating fuzzy logic with neural networks, this method facilitates a deeper understanding and design of the underlying mechanisms in BCI systems.
    
    \item \textbf{Practical Usability Improvements:} Our study demonstrates superior performance in handling new subject applications without the necessity for retraining or re-calibration. This significantly enhances the model’s practical usability and effectiveness in diverse real-world environments.

\end{enumerate}

\section{Methods and materials}

\subsection{Explanation of SSVEP Principles and Stimulus Frequency Modulation}

The principle behind SSVEP can be understood as a response of the sensory cortex to visual stimuli presented at specific frequencies, such as flicker \cite{muller2005steady} or other reversal patterns \cite{waytowich2016optimization}. This interaction results in an oscillatory brain response at both the stimulus frequency \(f_s\) and its harmonic frequencies \(kf_s\) (where \(k\) is a positive integer) \cite{moratti2007neural}. 

The most commonly used stimulus is flicker, whose chrominance value can be modulated sinusoidally to achieve a fixed frequency change:

\begin{align}
C(t) = \begin{bmatrix} 255 \\ 255 \\ 255 \end{bmatrix} \times \left( \frac{1 + \sin(2\pi f_s t + \phi)}{2} \right)
\end{align}

Here, \(C(t)\) represents the chrominance value at time \(t\), \(f_s\) is the frequency of the visual stimuli which can also be defined as \(y\) in the prediction task, \(\phi\) is the phase shift, and \(n\) denotes the number of target frequencies corresponding to \(n\) stimuli. By decoding the EEG frequency response of the subject, one can infer the target \(f_k\) that the subject is focusing on, thereby revealing their intentions.

By watching the flicker and recording the EEG signal from the occipital cortex, the ideal recorded brain response \(x(t)\) can be expressed as \cite{chenHighspeedSpellingNoninvasive2015a}:

\begin{align}
x(t) = \sum_{k=1}^{n} A_k \sin(2\pi k f_s t + \theta_k)
\end{align}

where \(A_k\) is the amplitude of the response at each harmonic \(k\) and \(\theta_k\) represents the phase associated with each harmonic frequency. This formulation highlights how the brain responds to the specific frequencies of visual stimuli, allowing for effective communication of the subject's focus.

\subsection{Task Definition and Data Structure}

We explore the domain of SSVEP tasks, incorporating data from \(N\) subjects to employ transfer learning techniques. The objective is to pretrain a model that adapts to new subjects under a zero-shot learning framework. Let \(\mathcal{S} = \{\mathcal{S}_1, \mathcal{S}_2, \ldots, \mathcal{S}_N\}\) represent the source domains, with each \(\mathcal{S}_n\) comprising pairs \(\{(x_{\mathcal{S}_n}^i, y_{\mathcal{S}_n}^i) \mid x_{\mathcal{S}_n}^i \in X_n, y_{\mathcal{S}_n}^i \in Y\}_{i=1}^{m_n}\). The target domain \(\mathcal{T}\), which consists of unlabeled samples \(\{x_{\mathcal{T}}^j \in X_{\mathcal{T}}\}_{j=1}^{m_T}\), aims to adapt using the learned knowledge from \(\mathcal{S}\) to predict labels \(\{y_{\mathcal{T}}^j \in Y\}_{j=1}^{m_T}\) effectively, achieving zero-shot learning.

\subsection{The proposed iFuzzyTL}

\begin{figure*}
    \centering
    \includegraphics[width=1\linewidth]{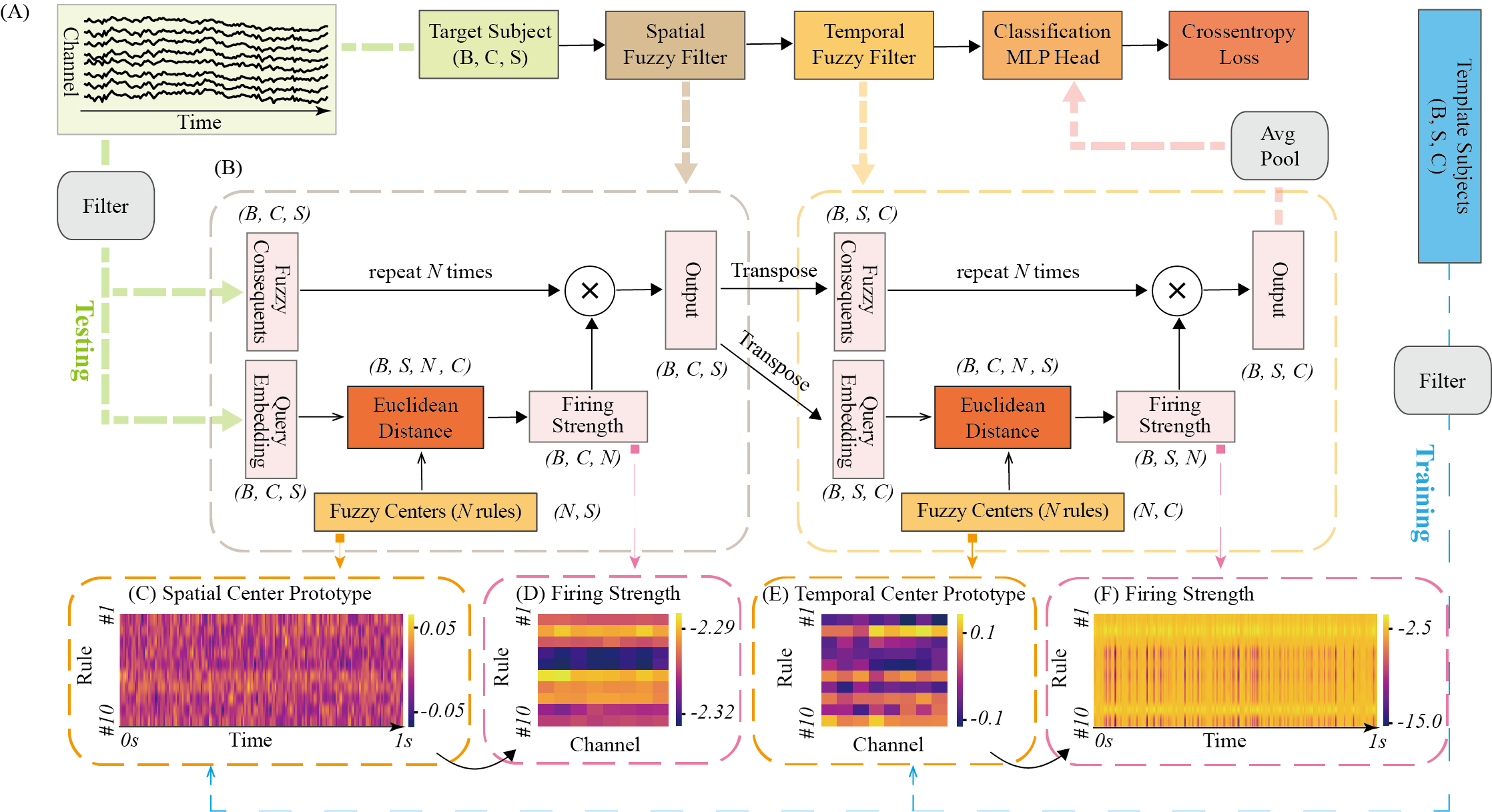}
    \caption{Illustration of the architecture for predicting target frequencies in an SSVEP task using the proposed iFuzzyTL model. 
    \textbf{(A)} Main structure of the iFuzzyTL model.
    \textbf{(B)} Design of the spatial and temporal fuzzy filters.
    \textbf{(C)} Detection of the center using the spatial fuzzy filter.
    \textbf{(D)} Firing strength of a demonstration sample to show the learned neural pattern as identified by the spatial fuzzy filter.
    \textbf{(E)} Detection of the center using the temporal fuzzy filter.
    \textbf{(F)} Firing strength of a demonstration sample to show the learned neural pattern as identified by the temporal fuzzy filter.
    }
    \label{fig:ModelArch}
\end{figure*}

\subsubsection{Proposition 1: Fuzzy Attention Layer as an Adaptive Filter in Signal Processing}
\label{prop:signal_filter}

Consider an input signal \( x(t) \) processed through an adaptive filter. The output \( Y(t) \) at time \( t \) is modeled as an adaptive linear combiner (ALC):

\begin{equation}
Y(t) = W_t^T \cdot x(t)
\end{equation}

where \( x(t) \) denotes the input feature vector at time \( t \), and \( W_t \) represents the adaptive weights. These weights are derived from a TSK fuzzy model, where the firing strength of a fuzzy rule is given by:

\begin{align}
\overline{f_r}(x(t)) 
    &= \frac{\mu_r(x(t))}{\sum_{i=1}^R \mu_i(x(t))} \nonumber \\
    &= \frac{\exp\left(-\sum_{d=1}^D \frac{(x_{t,d} - m_{r,d})^2}{2\sigma_{r,d}^2}\right)}{\sum_{i=1}^R \exp\left(-\sum_{d=1}^D \frac{(x_{t,d} - m_{i,d})^2}{2\sigma_{i,d}^2}\right)} \nonumber \\
    &= \mathrm{softmax}_{i,r}\left(-\sum_{d=1}^D \frac{(x_{t,d} - m_{r,d})^2}{2\sigma_{r,d}^2}\right) \nonumber\\
\label{eq:fuzzy-atten_time}
\end{align}

Here, \( r \) indexes the fuzzy rules, \( m_{r,d} \) represents the centers, and \( \sigma_{r,d} \) denotes the widths of the fuzzy sets for each rule dimension. Both parameters \( m_{r,d} \) and \( \sigma_{r,d} \) are learnable and evolve during training. By incorporating this fuzzy attention mechanism, we assign the adaptive weights as \( W_t^T = \overline{f_r}(x(t)) \), and the filter's output becomes:

\begin{equation}
    Y(t) = \overline{f_r}(x(t)) \cdot W_r^V x(t)
\label{eq:filter_design_time}
\end{equation}
where the projection is parameter matrices \( W_r^V \) for rule \( r \). This formulation enables the filter to adaptively modulate the importance of different features of \( x(t) \) based on their alignment with the fuzzy rule centers. The fuzzy attention mechanism dynamically adjusts the attention weights in response to the proximity of the input features to the fuzzy set centers, effectively allowing the filter to highlight or suppress certain signal features according to their fuzzy membership values.

\subsubsection{Proposition 2: Fuzzy Attention Layer as a Spatial Filter for Channel Selection}

Here, the Fuzzy Attention Layer acts as a spatial filter across different signal channels. The output \( Y \) is expressed as:

\begin{equation}
Y = W_c^T \cdot X
\end{equation}

where \( X \) represents the input feature matrix across channels, and \( W_c \) is the adaptive weight vector corresponding to each channel. The adaptive weights \( W_c \) are derived similarly to the time dimension in Proposition \ref{prop:signal_filter}, but now they focus on channel selection. The firing strength of a fuzzy rule for each channel \( c \) is given by:

\begin{align}
\overline{f_r}(X_c) = \mathrm{softmax}_{i,r}\left(-\sum_{d=1}^D \frac{(X_{c,d} - m_{r,d})^2}{2\sigma_{r,d}^2}\right)
\label{eq:fuzzy-atten_ch}
\end{align}

Thus, the spatial filter’s output becomes:

\begin{equation}
    Y = \overline{f_r}(X_c) \cdot W_{r}^VX
\label{eq:filter_design_ch}
\end{equation}

where the projection is parameter matrices \( W_r^V \) for rule \( r \). This allows the Fuzzy Attention Layer to adaptively weigh and select channels based on their proximity to the fuzzy rule centers, enhancing the filter's ability to focus on the most relevant channels for signal processing.

\subsubsection{Proposition 3: Input Recovery in Single-Layer Linear Networks}
\label{proposition:2}

This subsection demonstrates that the original input of a single-layer linear transformation, referred to as a projector, can be reconstructed from its output, termed the query. This recovery is contingent on the condition that the transformation matrix \( W \) is non-singular. The invertibility of \( W \) thus ensures the feasibility of interpretability within the iFuzzyTL framework.

Consider the linear transformation defined by:
\begin{equation}
\label{equa: ll}
y = Wx + b
\end{equation}
where \( y \) represents the output query, \( x \) the original input, \( W \) the transformation matrix, and \( b \) the bias vector.

To retrieve \( x \) from \( y \), rearrange the above equation to:
\begin{equation}
Wx = y - b
\end{equation}

Given \( W \) is non-singular, the inversion of \( W \) is feasible, allowing for the calculation of \( x \) by:
\begin{equation}
x = W^{-1}(y - b)
\end{equation}

This illustrates that the original input \( x \) is retrievable directly from the output \( y \) when \( W \) is invertible.

For scenarios where \( W \) is singular or not a square matrix, the recovery of \( x \) employs the Moore-Penrose pseudoinverse \( W^{+} \):
\begin{equation}
x = W^{+}(y - b)
\end{equation}

The computation of \( W^{+} \) utilizes the Singular Value Decomposition (SVD) of \( W \):
\begin{equation}
W = U \Sigma V^T
\end{equation}
where \( U \) and \( V \) are orthogonal matrices, and \( \Sigma \) contains the singular values.

The pseudoinverse \( W^{+} \) is then:
\begin{equation}
W^{+} = V \Sigma^{+} U^T
\end{equation}
with \( \Sigma^{+} \) derived by inverting the non-zero singular values of \( \Sigma \) and taking the transpose.

\begin{algorithm}
\caption{Pseudocode of recovering Input $x$ from Linear Layer \ref{equa: ll}}
\label{alg:recover_input}
\begin{algorithmic}[1] 
\REQUIRE $linear\_layer$, $y$  
\STATE $W \gets linear\_layer.weight.data$
\STATE $b \gets linear\_layer.bias.data$
\STATE $W_{\text{pinv}} \gets \text{pinverse}(W)$  
\STATE $recovered\_x \gets W_{\text{pinv}} \cdot (y - b)^T$  
\RETURN $recovered\_x$  
\end{algorithmic}
\end{algorithm}

This approach (as shown in Algorithm \ref{alg:recover_input}) guarantees that if the dimensions of the input data and the output query match, the reconstructed input will correspond to the original input.

In conclusion, under the condition that \( W \) is either invertible or suitably approximated via its pseudoinverse, the reversibility of the input from the output in a single-layer linear model is effectively demonstrated.

\subsubsection{Fuzzy Attention for SSVEP Transfer Learning}

Our proposed model, iFuzzyTL, integrates three key components tailored for SSVEP signal processing: the spatial fuzzy filter, the temporal fuzzy filter, and the classification head. Both fuzzy filters leverage the concept of Fuzzy Attention as outlined in Equations \ref{eq:fuzzy-atten_time} and \ref{eq:fuzzy-atten_ch}, and follow the design as the adaptive filers in Equations \ref{eq:filter_design_time} and \ref{eq:filter_design_ch}. The filter module can be defined as:

\begin{equation}
    \text{Fuzzyfilter}(x) = \text{Softmax}\left(\frac{(Q(x) - m)^2}{2\sigma^2}\right)u
\end{equation}

Here, \( Q(x) \) represents the linear projection of the input sequence, \( m \) and \( \sigma \) are the Gaussian membership parameters, and \( u \) is the consequent of the fuzzy rule. This formulation shifts the traditional attention mechanism from dot product-based similarity to an L2 distance weighted by \( \sigma \), thereby producing an adaptive fuzzy filter effect. 


The model architecture is depicted in Figure \ref{fig:ModelArch}(A). The spatial fuzzy filter initially processes the data, considering channel-like centers \( (N, C) \), followed by a transpose operation. Subsequently, the temporal fuzzy filter applies, which adapts to signal-like centers \( (N, S) \), where $N$ is the number of rules. This dual filtering strategy enables the model to encode both spatial and temporal dimensions of the SSVEP signals effectively. The structure of the spatial and temporal fuzzy filters are shown in Figure \ref{fig:ModelArch}(B).

The classification head consists of a 2-layer Multi-Layer Perceptron (MLP) model with ReLU (Rectified Linear Unit) activation and a dropout rate of 0.3 during training. The number of output nodes in the classification head corresponds to the number of labels.

The primary goal is to classify the SSVEP target frequencies accurately. We employ a multiclass cross-entropy loss function for this purpose, defined as:

\begin{equation}
    \text{loss}(y_{o,c}, p_{o,c}) = -\sum_{c=1}^{M} y_{o,c} \log(p_{o,c})
\end{equation}

Where \( y_{o,c} \) denotes the true label, \( p_{o,c} \) represents the predicted probability for class \( c \), and \( M \) is the total number of classes or target frequencies in the classification schema. This loss function quantifies the discrepancy between predicted probabilities and the actual class labels, facilitating practical model training to recognize SSVEP frequencies.

\subsection{Evaluation Metrics}

To evaluate the performance of each method, we used two primary metrics: classification accuracy and ITR. Classification accuracy is defined as the ratio of correctly classified samples to the total number of test samples. 

The ITR, measured in bits per minute (bits/min), quantifies the speed and accuracy of a brain-computer interface and is computed as follows \citep{WOLPAW2002767}:

\begin{equation}
    \text{ITR} = \frac{60}{T} \left[ \log_2 N + P \log_2 P + (1 - P) \log_2 \frac{1 - P}{N - 1} \right],
\end{equation}

where $T$ is the average time required for each selection operation, $N$ represents the number of possible classes, and $P$ is the classification accuracy. Following previous studies \cite{chenHighspeedSpellingNoninvasive2015a, nakanishi2017enhancing}, an additional 0.5 s was included in $T$ to account for gaze shift time. For example, if the data length is 1 s, $T$ is set to 1.5 s for the ITR calculation using the formula above.

In this study, we focused on zero-shot inter-subject classification experiments. We employed the leave-one-out cross-validation method, where the data from one subject was used as the test set while the data from all other subjects formed the training set, as shown in Figure \ref{fig:TLdemo}(C) and \ref{fig:ModelArch}(A). This process was repeated until each subject had been used as the test subject once, ensuring a complete evaluation. 

The baseline model and dataset description are in Supplementary Sections 1.1 and 1.2, respectively. 

\section{Results}

The average classification accuracies and ITR of the seven methods on the 12JFPM dataset are presented in Table \ref{tab:acc_12JFPM} and Supplementary Table I, respectively. Data lengths range from 0.5 s to 1.2 s in 0.1 s intervals, with additional lengths of 1.5 s and 2 s included to evaluate model performance over extended durations. The results indicate that iFuzzyTL consistently outperforms other baseline methods, achieving the highest average classification accuracies and ITR for data lengths under 1.2 s. However, for lengths of 1.5 s and 2 s, CCNN demonstrated superior performance compared to iFuzzyTL.

For classification accuracies, a two-way repeated measures ANOVA (rm-ANOVA) revealed significant main effects of data length ($F(9, 81) = 183.06$, $p < 0.001$) and method ($F(5, 45) = 30.56$, $p < 0.001$), as well as a significant interaction effect between them ($F(45, 405) = 21.66$, $p < 0.001$). Paired t-tests were conducted at each data length to compare iFuzzyTL with baseline methods, with statistical results summarized in Table \ref{tab:acc_12JFPM}. iFuzzyTL showed significant improvements over all methods except CCNN. It significantly outperformed CCNN for data lengths ranging from 0.6 s to 0.8 s ($p < 0.05$). 

Regarding ITR, the rm-ANOVA also indicated significant main effects of data length ($F(9, 81) = 6.88$, $p < 0.001$) and method ($F(5, 45) = 31.31$, $p < 0.001$), along with a significant interaction effect between them ($F(45, 405) = 11.28$, $p < 0.001$). Paired t-tests revealed that iFuzzyTL significantly outperformed all baseline methods except for CCNN, with significant differences observed at 0.7 s only. The detailed statistical results are presented in Supplementary Table I.

The average classification accuracies and ITR for the seven methods on the eldBETA dataset are shown in Table \ref{tab:acc_eldBETA} and Supplementary Table II, respectively. Data lengths span from 0.5 s to 1.2 s in 0.1 s intervals, with additional evaluations at 1.5 s and 2 s. The results again demonstrate that iFuzzyTL achieves superior performance across most data lengths, especially for durations under 0.7 s.

The two-way rm-ANOVA for classification accuracies revealed significant main effects of data length ($F(9, 891) = 371.72$, $p < 0.001$) and method ($F(5, 495) = 91.08$, $p < 0.001$), with a significant interaction effect ($F(45, 4455) = 27.41$, $p < 0.001$). Paired t-tests further indicated significant differences between iFuzzyTL and other methods, except CCNN. iFuzzyTL significantly outperformed CCNN at 0.5 s, 0.7 s, and 1.1 s ($p < 0.05$). For ITR, significant main effects of data length ($F(9, 891) = 37.88$, $p < 0.001$) and method ($F(5, 495) = 134.78$, $p < 0.001$), as well as a significant interaction effect ($F(45, 4455) = 49.36$, $p < 0.001$) were found. Paired t-tests showed that iFuzzyTL significantly outperformed all baseline methods, except CCNN at 0.7 s, 1.1 s, 1.5 s, and 2 s ($p < 0.05$). The detailed statistical results are presented in Supplementary Table II.

The average classification accuracies and ITR on the Benchmark dataset are reported in Table \ref{tab:acc_benchmark} and Supplementary Table III, respectively. Data lengths range from 0.5 s to 1.2 s, with additional evaluations at 1.5 s and 2 s. The results indicate that iFuzzyTL achieved the best performance at 0.5 s and for durations longer than 0.9 s ($p < 0.05$).

For classification accuracies, the two-way rm-ANOVA showed significant main effects of data length ($F(9, 306) = 393.52$, $p < 0.001$) and method ($F(5, 170) = 30.69$, $p < 0.001$), and a significant interaction effect ($F(45, 1530) = 29.20$, $p < 0.001$). Paired t-tests indicated significant differences between iFuzzyTL and other methods except CCNN. Regarding ITR, significant main effects of data length ($F(9, 306) = 26.01$, $p < 0.001$) and method ($F(5, 170) = 55.89$, $p < 0.001$), as well as a significant interaction effect ($F(45, 1530) = 36.39$, $p < 0.001$), were observed. The detailed statistical results are summarized in Supplementary Table III.

\begin{table*}[hp]
\centering
\captionsetup{width=\textwidth} 
\caption{Average accuracies (\%) across subjects for six methods at different data lengths on Dataset 12JFPM. Asterisks indicate significant differences between iFuzzyTL and the other methods, as assessed by paired t-tests (*$p < 0.05$, **$p < 0.01$, ***$p < 0.001$).}
\label{tab:acc_12JFPM}
\fontsize{5}{6}\selectfont
\setlength{\tabcolsep}{0.pt} 
\begin{tabular}{@{}cllllllllll@{}}
\toprule
Model & 0.5s & 0.6s & 0.7s & 0.8s & 0.9s & 1.0s & 1.1s & 1.2s & 1.5s & 2.0s \\ \midrule
TRCA & 18.33\% (6.32\%) *** & 22.94\% (9.16\%) *** & 30.89\% (12.52\%) *** & 36.61\% (15.98\%) *** & 45.67\% (18.27\%) *** & 52.94\% (22.06\%) *** & 57.61\% (22.83\%) *** & 62.11\% (24.00\%) *** & 74.39\% (23.63\%) ** & 84.39\% (19.72\%) * \\
eCCA & 54.94\% (21.86\%) * & 61.11\% (25.97\%) * & 64.39\% (27.22\%) ** & 67.78\% (28.34\%) * & 73.11\% (27.47\%) * & 75.56\% (26.35\%) * & 77.44\% (26.92\%) * & 79.78\% (24.80\%) * & 85.28\% (21.88\%) & 88.06\% (20.34\%) \\
CCNN & 64.61\% (22.29\%) & 66.56\% (23.43\%) *** & 72.33\% (23.23\%) * & 77.36\% (15.20\%) *** & 82.50\% (18.60\%) & 84.00\% (18.09\%) & 86.11\% (14.72\%) *** & 88.83\% (13.54\%) & 92.39\% (10.46\%) & 95.44\% (7.86\%) \\
EEGNET & 50.11\% (17.58\%) ** & 56.00\% (23.12\%) ** & 62.94\% (21.36\%) *** & 66.44\% (23.31\%) *** & 70.28\% (23.60\%) ** & 75.11\% (22.46\%) ** & 77.61\% (21.88\%) * & 79.72\% (22.35\%) * & 87.28\% (15.46\%) & 90.44\% (13.17\%) \\
SCCA\_qr & 54.83\% (21.93\%) * & 61.44\% (24.54\%) * & 66.89\% (26.25\%) ** & 72.39\% (27.99\%) * & 76.44\% (27.26\%) * & 78.83\% (25.84\%) * & 81.72\% (25.07\%) & 84.17\% (22.66\%) & 88.78\% (16.72\%) & 93.44\% (12.85\%) \\
iFuzzyTL & 67.92\% (12.59\%) & 75.29\% (15.28\%) & 81.91\% (16.42\%) & 84.76\% (15.36\%) & 88.64\% (16.10\%) & 89.70\% (14.89\%) & 90.22\% (13.99\%) & 90.14\% (13.88\%) & 91.97\% (15.42\%) & 92.41\% (14.26\%) \\
\bottomrule
\end{tabular}
\end{table*}

\begin{table*}[hp]
\centering
\captionsetup{width=\textwidth} 
\caption{Average accuracies (\%) across subjects for six methods at different data lengths on Dataset eldBETA. Asterisks indicate significant differences between iFuzzyTL and the other methods, as assessed by paired t-tests (*$p < 0.05$, **$p < 0.01$, ***$p < 0.001$).}
\label{tab:acc_eldBETA}
\fontsize{5}{6}\selectfont
\setlength{\tabcolsep}{0.pt} 
\begin{tabular}{@{}cllllllllll@{}}
\toprule
Model & 0.5s & 0.6s & 0.7s & 0.8s & 0.9s & 1.0s & 1.1s & 1.2s & 1.5s & 2.0s \\ \midrule
TRCA & 42.05\% (19.30\%) *** & 44.17\% (20.15\%) *** & 45.75\% (20.98\%) *** & 48.40\% (21.45\%) *** & 51.14\% (22.69\%) *** & 52.54\% (23.20\%) *** & 55.17\% (23.26\%) *** & 56.51\% (23.68\%) *** & 60.48\% (24.23\%) *** & 61.86\% (24.63\%) *** \\
eCCA & 46.60\% (19.35\%) *** & 51.16\% (20.26\%) *** & 55.52\% (21.57\%) *** & 58.73\% (22.32\%) *** & 62.60\% (22.47\%) *** & 65.54\% (22.62\%) *** & 68.16\% (22.32\%) *** & 70.54\% (21.74\%) *** & 75.30\% (20.66\%) *** & 80.14\% (19.25\%) *** \\
CCNN & 62.29\% (21.09\%) ** & 65.05\% (21.14\%) & 68.63\% (21.52\%) *** & 70.90\% (21.52\%) & 73.05\% (21.62\%) & 74.95\% (21.04\%) & 76.30\% (20.70\%) ** & 77.63\% (20.47\%) & 81.14\% (19.04\%) ** & 83.98\% (17.97\%) ** \\
EEGNET & 57.51\% (21.14\%) *** & 59.46\% (22.38\%) *** & 62.00\% (23.06\%) *** & 64.37\% (22.63\%) *** & 65.97\% (23.22\%) *** & 67.25\% (23.43\%) *** & 68.65\% (23.20\%) *** & 69.24\% (23.76\%) *** & 71.76\% (23.29\%) *** & 74.13\% (22.87\%) *** \\
SCCA\_qr & 41.70\% (19.48\%) *** & 49.22\% (21.13\%) *** & 53.84\% (22.04\%) *** & 57.92\% (22.37\%) *** & 61.38\% (22.69\%) *** & 64.75\% (22.68\%) *** & 67.35\% (22.56\%) *** & 69.30\% (22.63\%) *** & 74.62\% (21.52\%) *** & 81.00\% (18.90\%) *** \\
iFuzzyTL & 66.48\% (15.09\%) & 66.85\% (15.46\%) & 74.02\% (16.91\%) & 71.45\% (17.40\%) & 74.00\% (17.54\%) & 76.50\% (17.19\%) & 80.17\% (17.34\%) & 78.82\% (18.05\%) & 84.16\% (16.35\%) & 86.70\% (15.67\%) \\
\bottomrule
\end{tabular}
\end{table*}

\begin{table*}[hp]
\centering
\captionsetup{width=\textwidth} 
\caption{Average accuracies (\%) across subjects for six methods at different data lengths on Dataset Benchmark. Asterisks indicate significant differences between iFuzzyTL and the other methods, as assessed by paired t-tests (*$p < 0.05$, **$p < 0.01$, ***$p < 0.001$).}
\label{tab:acc_benchmark}
\fontsize{5}{6}\selectfont
\setlength{\tabcolsep}{0.pt} 
\begin{tabular}{@{}cllllllllll@{}}
\toprule
Model & 0.5s & 0.6s & 0.7s & 0.8s & 0.9s & 1.0s & 1.1s & 1.2s & 1.5s & 2.0s \\ \midrule
TRCA & 33.39\% (21.44\%) *** & 38.80\% (24.15\%) *** & 44.65\% (25.67\%) *** & 50.85\% (26.54\%) *** & 55.52\% (27.08\%) *** & 58.86\% (27.60\%) *** & 60.77\% (27.52\%) *** & 62.07\% (27.49\%) *** & 66.76\% (26.54\%) *** & 75.35\% (23.81\%) *** \\
eCCA & 39.17\% (18.58\%) *** & 48.69\% (21.28\%) *** & 59.64\% (21.41\%) ** & 67.04\% (21.16\%) *** & 72.74\% (20.13\%) *** & 76.42\% (19.62\%) ** & 78.63\% (19.46\%) *** & 81.95\% (17.97\%) *** & 87.25\% (14.70\%) ** & 93.42\% (8.51\%) * \\
CCNN & 52.49\% (18.50\%) *** & 62.37\% (19.86\%) & 68.92\% (20.63\%) & 73.58\% (19.94\%) & 76.27\% (19.85\%) ** & 78.77\% (19.47\%) ** & 81.44\% (18.96\%) *** & 82.75\% (18.80\%) *** & 87.79\% (16.97\%) * & 91.36\% (14.41\%) * \\
EEGNET & 39.86\% (18.36\%) *** & 46.32\% (20.73\%) *** & 54.95\% (21.62\%) *** & 60.62\% (22.47\%) *** & 65.08\% (21.90\%) *** & 66.44\% (22.55\%) *** & 68.32\% (21.88\%) *** & 70.58\% (22.07\%) *** & 74.50\% (21.74\%) *** & 81.92\% (20.13\%) *** \\
SCCA\_qr & 19.04\% (10.83\%) *** & 27.88\% (14.59\%) *** & 38.07\% (18.44\%) *** & 48.24\% (20.47\%) *** & 55.99\% (21.20\%) *** & 62.49\% (21.12\%) *** & 67.67\% (20.70\%) *** & 72.37\% (20.86\%) *** & 80.70\% (18.53\%) *** & 88.96\% (13.95\%) *** \\
iFuzzyTL & 62.22\% (7.28\%) & 65.33\% (9.86\%) & 69.78\% (12.53\%) & 77.88\% (13.54\%) & 83.39\% (13.97\%) & 85.81\% (15.16\%) & 89.34\% (12.74\%) & 90.56\% (11.60\%) & 93.18\% (10.01\%) & 95.97\% (5.61\%) \\
\bottomrule
\end{tabular}
\end{table*}

\section{Real-Time Feasibility Evaluation}
To evaluate the feasibility of the proposed model in real-world applications, we conducted an online experiment consisting of a data collection session and an online test session.

\begin{figure*}[htp]
    \centering
    \includegraphics[width=1\linewidth]{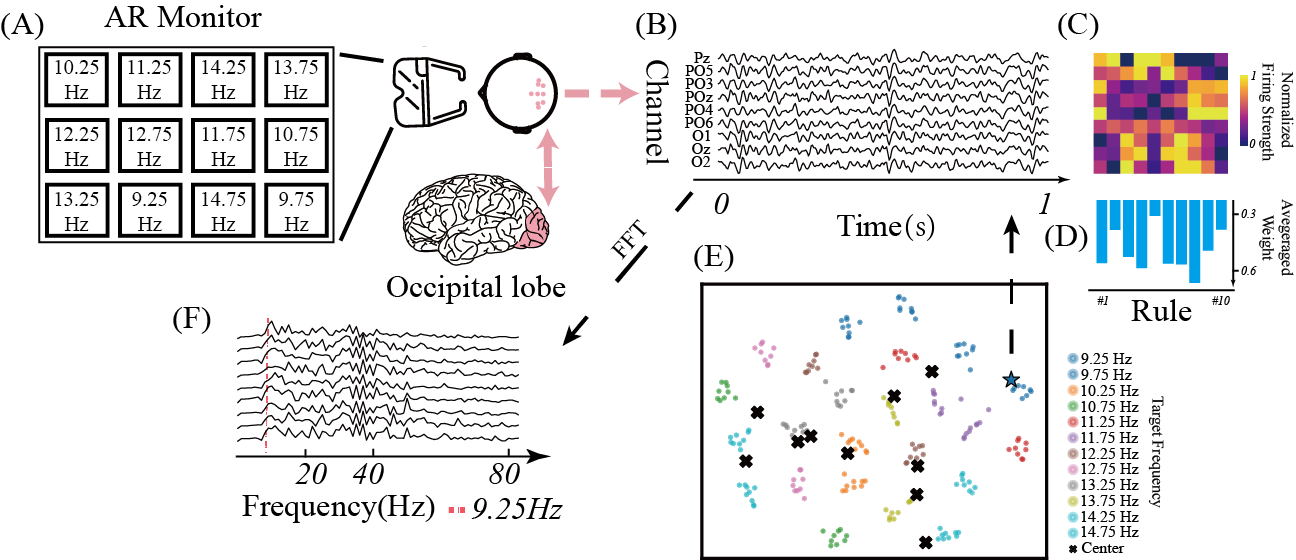}
    \caption{Detailed illustration of a real-time SSVEP experiment setup.
    \textbf{(A)} Demonstration setup for the experiment. The EEG channels are located in the occipital lobe. 
    \textbf{(B)} Example of a filtered EEG signal used in the demo.
    \textbf{(C)} Min-max normalized firing strength across the rules.
    \textbf{(D)} Averaged weight distribution among the rules (without normalization).
    \textbf{(E)} Data distribution following the application of the spatial filter.
    \textbf{(F)} Fourier Transform results of the demo EEG signal.}
    \label{fig:online}
\end{figure*}

\subsection{Experiment Design}
Following the methodology of the 12JFPM dataset, twelve cube flickers were selected as stimuli, with frequencies ranging from 9.25 Hz to 14.75 Hz in 0.5 Hz increments, as shown in Figure \ref{fig:online}(A). The cube flickers alternated between black and white, with sinusoidal coding used to control their chrominance values. Unlike the previously described dataset, the stimuli in this experiment were implemented using Unity 3D and presented in an augmented reality (AR) environment built with a Hololens headset. Specifically, the twelve flickers were arranged in a $3 \times 4$ grid layout at the center of a real-world scenario. Each flicker had dimensions of $472 \times 472$ pixels ($2.3^\circ \times 2.1^\circ$ in visual angle), with a distance of $150$ pixels between adjacent flickers.

\subsection{Participants and Data Acquisition}
Six subjects, comprising five males and one female aged between 22 and 27 years (\(\mu = 24.5\), \(\sigma = 1.64\)), participated in this study. All participants were mentally healthy and possessed normal or corrected-to-normal vision. Prior to the experiment, written ethical consent, approved by the University of Technology Sydney's Ethical Committee (Grant number: UTS HREC REF No. ETH20-5371), was obtained from each participant. EEG signals were recorded using a 64-electrode Neuroscan system, supplemented by a pair of mastoid reference electrodes and a Synamps2 amplifier (Compumedics Neuroscan, Charlotte, NC, USA). The sampling rate for recording was set at 250 Hz.

\subsection{Procedure for Training Data Collection}
A small dataset was collected for model training. For each subject, five rounds of data collection were conducted. In each round, one of the twelve flickers was highlighted in red as a cue for 1000 ms. After the cue disappeared, all flickers began flashing, and the subject was instructed to focus on the cued flicker. The flashing lasted for 5000 ms, followed by a 3000 ms rest period. A new flicker, which had not yet been cued, was then selected as the target, and a new trial began. Each round consisted of twelve trials, with all flickers being traversed once as targets. A total of 60 samples were collected per subject.

\subsection{Data Preprocessing}
For classification, signals from nine channels—Pz, PO3, PO4, PO5, PO6, POz, O1, O2, and Oz—were utilized. To ensure the feasibility of online testing, the preprocessing was kept minimal. A bandpass filter with a frequency range of 8 Hz to 90 Hz was applied to remove artifacts. The filtered data was then fed into the model for classification.

\subsection{Online Test}
During the online session, a ‘leave-one-out’ strategy was employed for evaluation. For each subject, data from the remaining five subjects were used for model training. The pre-trained model was then integrated into the online system. In each online trial, a cue was presented, and the subject focused on the target flashing flicker. At the start of the flashing, a ‘start’ marker was sent to the terminal via UDP, triggering the online system to record task-related EEG data. When the flashing ended, an ‘end’ marker was sent, and the EEG data was segmented. The segmented EEG was then input into the pre-trained model, and the decoding result was recorded. Each subject completed three rounds of the online experiment, resulting in 36 trials per subject. The accuracy and ITR result of the 1s real-time experiment is 90.12\% $\pm$ 2.27\% and 164.93 $\pm$ 9.11, respectively.

\section{Discussion}

In the iFuzzyTL framework, the \textit{center} plays a key role by encapsulating domain knowledge derived from source domains. It acts as a general \textit{template}, effectively capturing the essence of the source domain characteristics. By computing the distance between this learned center and incoming data points, the model robustly leverages the underlying domain knowledge to make informed decisions. This mechanism facilitates robust decision-making and significantly enhances the model's transferability across different SSVEP tasks. Incorporating the center as a template proves advantageous for SSVEP applications, where the ability to generalize across varying conditions and subjects is crucial. Consequently, iFuzzyTL offers an improved approach to handling the inherent variability in SSVEP signals, ensuring higher performance and reliability in real-world scenarios.

\subsection{Sample-wised Interpretability Analysis}

\begin{figure*}[!htp]
    \centering
    \includegraphics[width=1\linewidth]{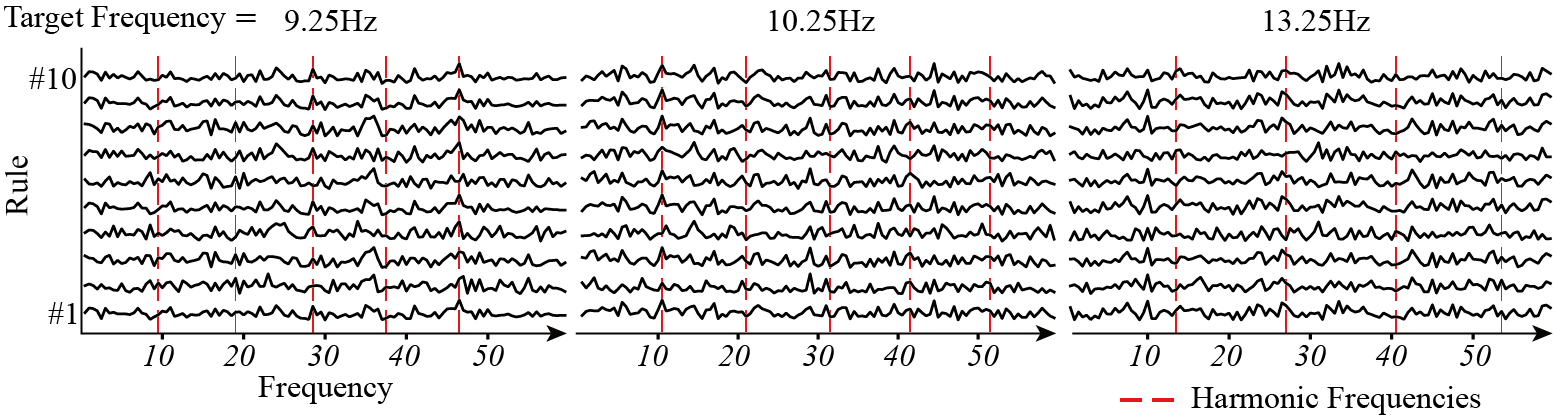}
    \caption{FFT features that triggered each fuzzy rule across different target frequencies, illustrating the identification of harmonic peaks.}
    \label{fig:case_fs}
\end{figure*}

\subsubsection{Demo Analysis of SSVEP Target Frequency Identification Using iFuzzyTL Model}
To provide an intuitive understanding of how the iFuzzyTL model identifies the SSVEP target frequency, we present a demo sample from the best-performing subject (S8) in the 12JFPM dataset, with a target frequency of 9.25 Hz, as shown in Figure \ref{fig:ModelArch}(A). Figure \ref{fig:ModelArch}(C) illustrates that the spatial fuzzy filter's center pattern resembles the EEG signal, with distinct phases for each rule. In Figure \ref{fig:ModelArch}(D), the border firing strength indicates that the contributions of rules \#4 and \#5 for this sample are minimal, while channels 1 and 2 contribute significantly to rule \#6.

After applying the spatial filter, Figure \ref{fig:ModelArch}(E) demonstrates the center of the temporal fuzzy filter for subject S8, displaying the neural patterns captured across 10 separate rules. Particularly, rule \#2 shows a high contribution in most channels, whereas rules \#1 and \#2 exhibit the lowest contribution. Additionally, Figure \ref{fig:ModelArch}(F) reveals that the firing strength in the temporal fuzzy filter indicates a stable pattern for rules \#4-8 and \#10 in this sample.

\subsubsection{Analysis of Temporal Firing Strength Features in the Frequency Domain}

To further investigate how the fuzzy rules learns features in the time domain, we compute the FFT of the input feature that triggered the fuzzy rules in the proposed temporal filter. The results indicate that the firing strength exhibits peaks at harmonic frequencies in correct case, as demonstrated in Figure \ref{fig:case_fs}. 

In the left panel, corresponding to a target frequency of 9.25 Hz, a prominent peak is observed at 28 Hz. This shift is attributed to the relatively low sample rate, which causes the harmonic frequencies in the FFT to exhibit slight displacements. The middle panel illustrates a target frequency of 10.25 Hz, with peaks observed at both 10.5 Hz and 41.5 Hz. In contrast, the right panel presents a less favorable case with a target frequency of 13.25 Hz, where peaks near 27 Hz are detected across several rules, indicating that not all rules accurately capture the target frequency, although some rules remain effective.

As shown in Supplementary Figure 1, this case illustrates why the model failed to predict the SSVEP signal, with all peaks misaligned with the harmonic frequencies.

\subsection{Deeper Analysis of Correct and Incorrect Cases}
\begin{figure*}
    \centering
    \includegraphics[width=0.9\linewidth]{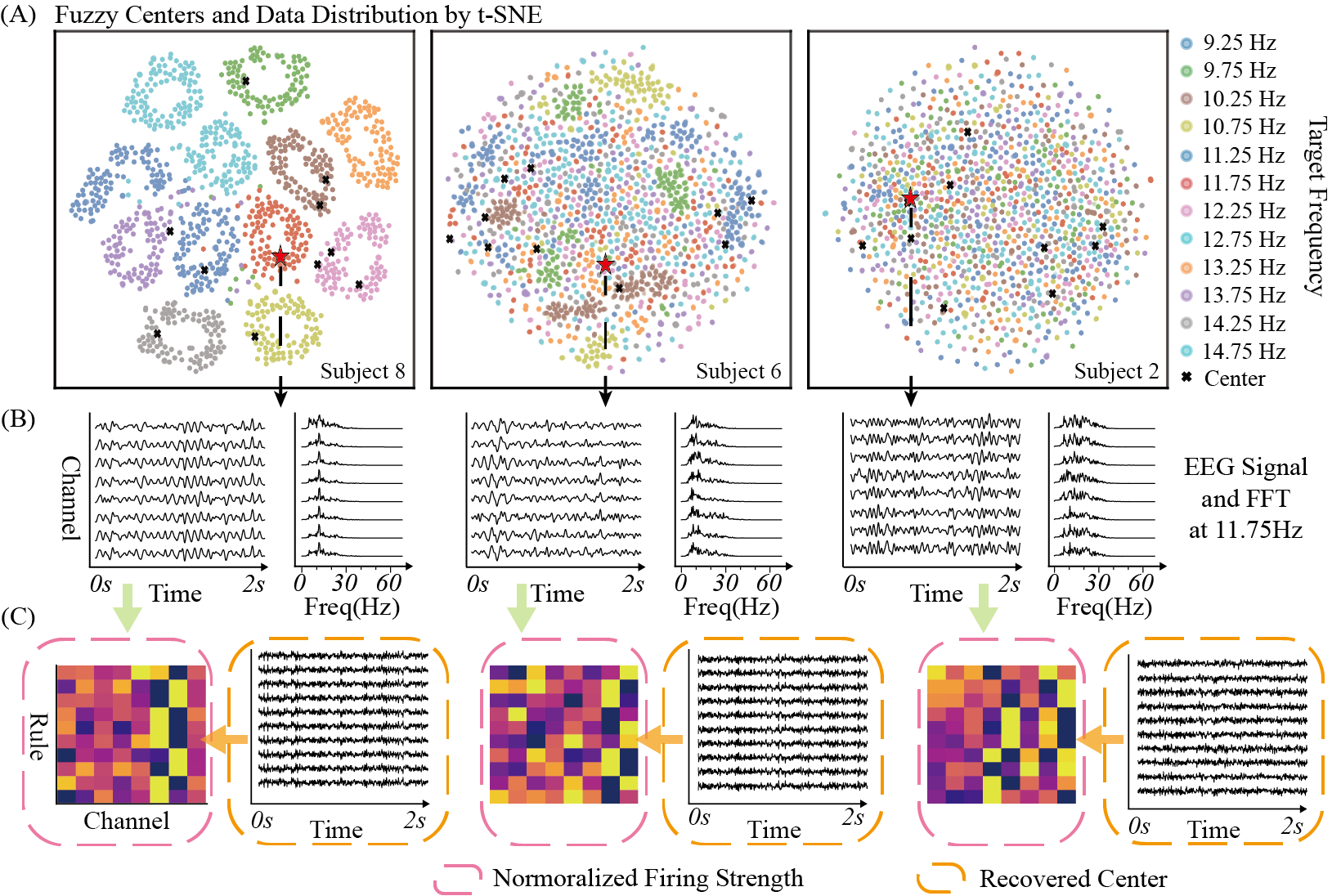}
    \caption{Visualization of demographic subjects from the 12JFPM dataset (2s) illustrating what iFuzzyTL learned. 
    \textbf{(A)} Data distribution post-application of the spatial filter, highlighting the position (Red Star) of a sample needing explanation at 11.75Hz.
    \textbf{(B)} Filtered EEG signals and their Fourier Transform to display the data characteristics.
    \textbf{(C)} Representation of firing strength and the center, using min-max normalization across the channel dimension to accentuate differences within one rule. The center is reconstructed from the query space to the raw EEG signal space as described in proposition \ref{proposition:2}.}
    \label{fig:SampleExplain}
\end{figure*}

For deeper understanding of correct and incorrect cases of iFuzzyTL model, we randomly select one sample from the same target frequency (11.75 Hz) from three subjects (S2, S6, and S8) who exhibited varying performance levels (input data length as $1$s: S2: $55.12\%$, S6: $96.95\%$, S8: $99.17\%$; input data length as $2$s: S2: $55.12\%$, S6: $100\%$, S8: $100\%$). Subject S8 demonstrated the best performance among all subjects in the 12JFPM dataset across all tested models, while subject S6 performed better than subject S2 but worse than subject S8. Subject S2 showed the worst performance across all models. Then, we compare how the iFuzzyTL model processes the EEG signal in three different-quality subjects.

As shown in Figure \ref{fig:SampleExplain}(A), the data distribution after applying the spatial filter for subject S8 is highly clustered and distinct, whereas subject S6 shows a less distinct clustering pattern and subject S2 exhibits the weakest clustering. For subject S8, the fuzzy centers are mostly located within the clusters corresponding to each target frequency, whereas for subjects S6 and S2, the center alignment is less clear. 

In Figure \ref{fig:SampleExplain}(C), The firing strength of the spatial fuzzy filter for the selected sample of subject S8 is more consistent across the rules, with Channels 6 and 7 significantly contributing to the decision-making process. As for subject S6, the contributions vary more across different rules, with Channels 7 and 8 being important overall, while Channels 2, 3, and 5 play key roles in specific rules (\#4, \#1\&2, and \#6\&10, respectively). In contrast, the firing strength of subject S2 has an unclear pattern. Channels 4-8 show major contributions across different rules. 

To better understand these differences, we refer to Figure \ref{fig:SampleExplain}(B). It is well-known that FFT features are crucial for SSVEP. In subject S8, the signal quality across all channels in this sample is high, enabling the spatial filter to effectively select the optimal channels for adaptive filtering, thereby enhancing model prediction accuracy. As for subject S6, the channel quality is moderate, as evidenced by multiple peaks in the FFT, which include both target and harmonic frequencies. Consequently, the filter relies on information from more channels to achieve satisfactory results and eventually make correct predictions. Conversely, subject S2 exhibits poor overall channel quality, with detectable peaks in Channels 6, 7, 8, 9, and 10 in the FFT though the peak is not clear. However, Channels 1, 2, and 3 display multiple unclear frequency peaks. This causes the spatial filter to apply different rules when selecting these channels. Thus, distinctive rules are created in a fragmented manner, and incorrect predictions are made in subject S2.

Similarly, in our real-time test, as shown in Figures \ref{fig:online}(B) (raw signal) and \ref{fig:online}(F) (FFT result), the FFT peaks of Channels Pz and PO5 are prominent, and their contribution in Rule \#4 is substantial. The data distribution for this scenario is also as clear as that of subject S8 in the 12JFPM dataset, as illustrated in Figure \ref{fig:online}(E). Particularly, during the application of the spatial filter, the fuzzy attention mechanism may not strictly filter channels. For example, as shown in Figure \ref{fig:online}(D), Rule \#8 has the highest average weight, yet it selects Channels PO3, POz, and PO4, which are not strongly indicated by FFT. This discrepancy may be attributed to the limited diversity of the small training set.

In summary, these findings suggest that iFuzzyTL can better learn the knowledge from the source domain and predict the result if the signal is clear; otherwise, iFuzzyTL can effectively select high-quality channels by the spatial filter through a combination of rules, thus enhancing the filtering of input signals, but may cause the incorrect predictions.

\subsection{Ablation Study}

The ablation study was meticulously designed to rigorously evaluate the influence of various parameters and modules of our proposed Fuzzy rule-based framework, iFuzzyTL, on its performance. This systematic examination helps uncover the contributions of individual components and configurations to the overall efficacy and operational dynamics of the model. All statistical comparisons were conducted using paired t-tests to assess the significance of differences.

\subsubsection{Filter Effect Analysis}
\label{sec:filtermodule}

\begin{figure}[htp]
    \centering
    \includegraphics[width=\linewidth]{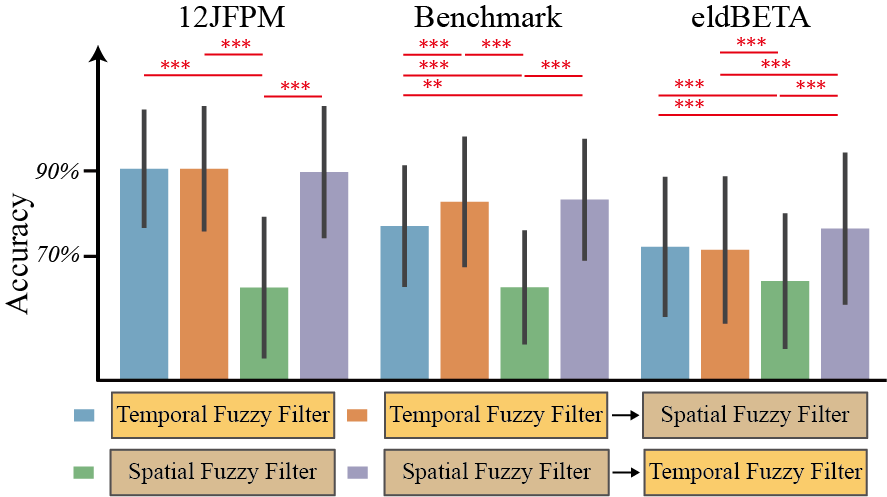}
    \caption{Variation in accuracy as a function of different configurations of two fuzzy filter modules across three datasets. This figure illustrates the significant impact of filter type and sequence on model performance. **$p < 0.01$, ***$p < 0.001$}
    \label{fig:module}
\end{figure}

In this study, we assessed the performance impact of various configurations and types of Fuzzy filters on signal processing. The primary motivation was to elucidate the relative importance of spatial versus temporal filtering and to determine the influence of their application sequence on the quality of the resultant data. The significance test and result are shown in Figure \ref{fig:module}.

\paragraph{Impact of Filter Application Order}
Our experimental setup tested two different sequences of applying filters to ascertain their influence on signal integrity and feature isolation. The first sequence involved applying a spatial filter followed by a temporal filter aimed at reducing spatial noise to enhance signal clarity before isolating temporal features. The alternative sequence started with a temporal filter intended to highlight temporal dynamics, followed by a spatial filter to refine the signal's spatial characteristics. Notably, significant differences were observed predominantly in the eldBETA dataset ($p < 0.05$), where the spatial filter followed by the temporal filter sequence exhibited superior performance compared to the reverse sequence.

\paragraph{Evaluating Single Filter Types}
The investigation also explored the effects of using each type of filter independently. The application of only a spatial filter was to evaluate the consequences of omitting temporal filtering, whereas using only a temporal filter was intended to assess whether spatial information alone could suffice for specific analytical tasks. The findings revealed that the exclusive use of a Spatial Filter significantly reduced performance across all datasets ($p < 0.05$). In contrast, employing only a temporal filter maintained relatively high performance, though it was slightly outperformed by the combined filter sequences in the Benchmark dataset.

After that, the comparison of the single filter with a combination of two filters is also tested. The results show that the bi-directional combination of two filters ($temporal filter \rightarrow spatial filter$ and reversed) is significantly better than the single spatial filter ($p < 0.05$) in three datasets. the Benchmark dataset shows a significant improvement from a single temporal filter to the bi-directional combination of two filters($p < 0.05$), and the eldBETA dataset shows a significant improvement from a single temporal filter to $spatial filter \rightarrow temporal filter$ combination of two filters($p < 0.05$).

The results underscore the indispensable nature of the Temporal Filter, whereas the spatial filter, though beneficial, proved less critical. The sequence of filter application did not significantly impact performance, except in certain datasets where the Spatial Filter followed by Temporal Filter configuration slightly outperformed others. These findings contrast traditional approaches such as CCA, which primarily relies on spatial filtering. Moreover, recent methodologies that integrate temporal filters, such as TRCA and ECCA, further validate the relevance of our results \cite{ou2023improving}.

\subsubsection{Number of Rule Effect Analysis}
\label{sec:NRules}
\begin{figure} [htp]
    \centering
    \includegraphics[width=1\linewidth]{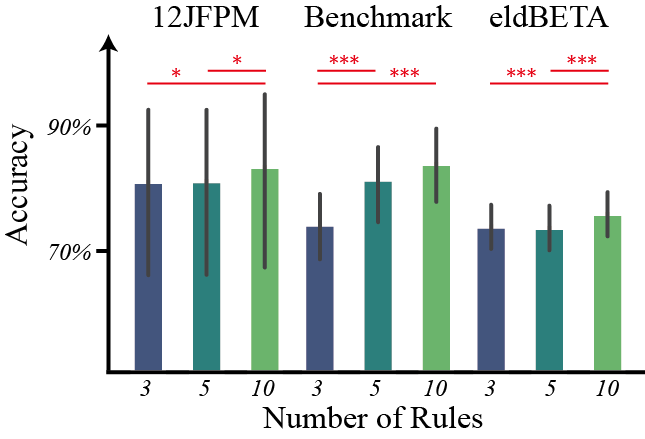}
    \caption{Accuracy variation as a function of the number of rules across three datasets, highlighting the impact of rule number on model performance. *$p < 0.05$, ***$p < 0.001$}
    \label{fig:NumberOfRules}
\end{figure}

In this ablation study, which examines the impact of the number of rules within our model, Figure \ref{fig:NumberOfRules} demonstrates the variability in accuracy across the 12JFPM, Benchmark, and eldBETA datasets with varying rule counts of 3, 5, and 10. Our results indicate that an increase from 3 to 10 rules generally enhances accuracy across all evaluated datasets. This suggests that a moderate increase in model complexity can positively affect the model's transfer learning capabilities. Particularly, the Benchmark dataset shows the most significant enhancements when the rule count is increased from 3 to 5 and subsequently from 5 to 10, with all transitions showing statistical significance ($p < 0.05$). Meanwhile, the 12JFPM and eldBETA datasets display significant improvements predominantly for transitions from 5 to 10 and 3 to 10 ($p < 0.05$). Summarily, Incorporating more rules extends the knowledge coverage, thereby increasing the capacity for domain adaptation, which is critical for effective transfer learning.

\subsubsection{FFT Features Versus Time-domain Features}

In this ablation study, we evaluate the influence of various input features on the model's performance across three datasets, comparing the effectiveness of FFT features against time-domain features. We structured the input data in two formats: FFT features represented as \((n_{\text{channel}}, n_{\text{freq}})\) and time-domain features represented as \((n_{\text{channel}}, n_{\text{timepoint}})\). The FFT feature is calculated based on a 1s time window.

\begin{figure}
    \centering
    \includegraphics[width=\linewidth]{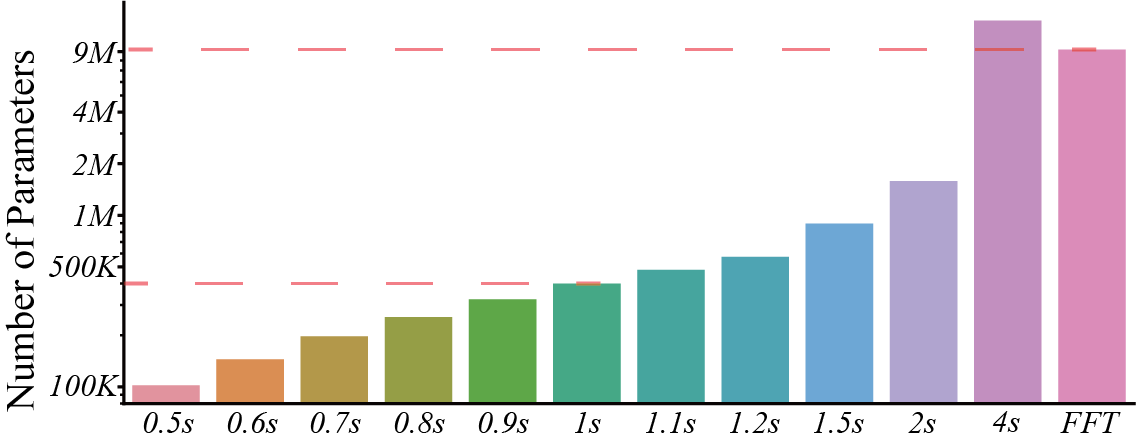}
    \caption{Number of parameters for models trained with different input features.}
    \label{fig:Nparameters}
\end{figure}

\begin{figure}
    \centering
    \includegraphics[width=1\linewidth]{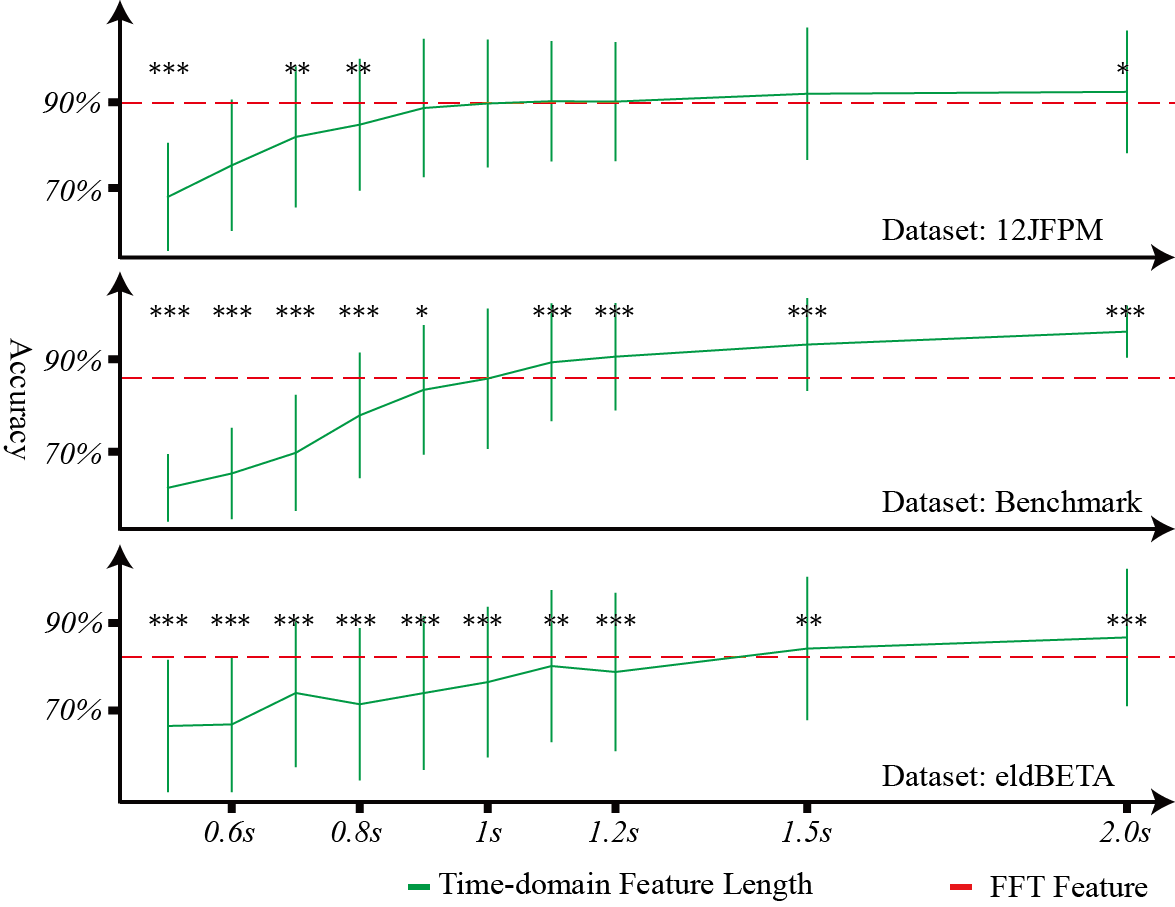}
    \caption{Comparison of FFT and time-domain features across different datasets. The red dashed line represents the performance using FFT features (calculated from 1-second data), while the green line shows the performance with varying lengths of time-domain features. Asterisks indicate statistically significant differences between the FFT and time-domain features: *$p < 0.05$, **$p < 0.01$, ***$p < 0.001$.}
    \label{fig:fft}
\end{figure}

As shown in Figure \ref{fig:fft}, our results indicate that the performance of FFT features is comparable to the 1-second time-domain features, except for the eldBETA dataset ($p > 0.05$). However, for all datasets, the 2-second time-domain features outperform FFT features ($p < 0.05$), suggesting that time-domain features are more effective at capturing relevant information from the signals. Figure \ref{fig:Nparameters} further illustrates that in the 12JFPM dataset, the model trained on FFT features has over 9 million parameters, while the model using 1-second time-domain features with a 256Hz sample rate requires only around 400,000 parameters. As the time window increases to 4 seconds, the model's parameter count rises to 13.7 million, surpassing that of the FFT-based model. This demonstrates that iFuzzyTL can achieve competitive performance with significantly fewer parameters when using time-domain features.

These findings highlight the capability of our proposed model, iFuzzyTL, to effectively extract essential information directly from raw signals without the need for additional feature extraction steps such as FFT. This approach not only simplifies the preprocessing pipeline but also enhances the model's ability to leverage the inherent temporal dynamics of the data, leading to improved performance while maintaining a relatively low parameter count. However, there is a trade-off between model complexity and performance in real-time applications. While time-domain features offer a significant performance advantage, especially in longer time windows, the choice of feature type should balance both parameter efficiency and accuracy based on the specific application requirements.


\subsection{Limitations}
Our study has identified several limitations with the iFuzzyTL model. Firstly, the model (rule count is 10) comprises approximately 400K parameters, which leads to longer training times, potentially limiting its efficiency in scenarios requiring quick deployment and can not automatically select the number of rules. Secondly, iFuzzyTL cannot be directly tested with a different set of devices if the electrode channel locations vary, as the model's performance is contingent on specific channel configurations. Lastly, the model does not support direct testing with different target frequencies without adjustments, which may restrict its application across diverse BCI setups where frequency variations are common. 

Future work can further investigate methods to reduce the parameter count while maintaining or enhancing performance, automatically decide the number of rules, explore adaptable channel configuration strategies for greater device compatibility, and develop frequency-independent processing techniques to accommodate varying target frequencies such as regression model. This will potentially broaden the applicability of iFuzzyTL across a wider range of BCI systems and real-world scenarios.

\section{Conclusion}
In this work, we introduce iFuzzyTL, a fuzzy logic-based attention mechanism for enhancing transfer learning in SSVEP BCI systems. iFuzzyTL significantly reduces the need for user-specific calibration, facilitating a plug-and-play experience that mimics human cognitive processes for improved interpretability and transparency. Our interdisciplinary approach integrates fuzzy logic with neural networks, enhancing model transferability and interpretability, crucial for zero-shot learning applications. Experimental results demonstrate that iFuzzyTL not only achieves superior recognition accuracy in zero-calibration scenarios but also surpasses real-time experiments, confirming its practical usability in diverse real-world environments without the need for retraining or re-calibration. This advancement addresses common challenges in deploying SSVEP-based BCIs, especially in dynamic settings where re-calibration is impractical, making iFuzzyTL a promising solution for high-performance, low-calibration, and easily implementable practical BCI applications.


\bibliographystyle{IEEEtran}
\bibliography{ref, ref_Beining}
\end{document}


\label{SupplementaryMaterial}
\maketitle
\section{Methods and materials}
\subsection{Dataset}
\label{sec:dataset}

The proposed transfer learning method is used to evaluate three public datasets. Each dataset is tailored to capture SSVEP responses under varying conditions and setups.

\subsubsection{12JFPM}
The first dataset, 12JFPM \cite{12JFPM}, was collected using a 27-inch LCD monitor to display 12 visual target stimuli. EEG signals were recorded using the BioSemi ActiveTwo system equipped with eight electrodes over the occipital region of the scalp. The stimuli frequencies ranged from 9.25 Hz to 14.75 Hz in increments of 0.5 Hz, with initial phases increasing in 0.5$\pi$ steps. The experimental setup included 15 rounds, each comprising 12 trials corresponding to the target stimuli. Participants were instructed to fixate on a randomly selected target for 4 seconds per trial. The EEG signals were sampled at 2048 Hz and then downsampled to 256 Hz for analysis.

\subsubsection{Benchmark}
The second dataset, Benchmark \cite{liu2020beta}, included 40 visual targets displayed on an LCD monitor, with frequencies ranging from 8 Hz to 15.8 Hz in 0.2 Hz increments and initial phases starting at zero, increasing by 0.5$\pi$ steps. A total of 35 participants, including eight experienced with SSVEP-based spellers, took part in six experimental rounds. Each trial lasted 6 seconds, encompassing a 0.5-second cue period and a 0.5-second blank period at the end. The effective SSVEP data were thus collected between 0.5 seconds and 5.5 seconds of each trial. Data acquisition was performed using a Synamps2 EEG system (Neuroscan, Inc.), with a sampling rate of 1000 Hz reduced to 250 Hz post-acquisition. A 50 Hz notch filter was applied to eliminate power frequency interference.

\subsubsection{eldBETA}
The third dataset, eldBETA \cite{liuEldBETALargeEldercareoriented2022}, was composed of data from 100 elderly subjects ranging from 63 to 81 years old. The dataset featured nine target frequencies from 8 Hz to 12 Hz in 0.5 Hz increments, with initial phases starting at zero and increasing by 0.5$\pi$ steps. Participants underwent seven rounds of experiments, each involving a cue period marked by a green-colored flicker for 4 seconds, followed by 5 seconds of gazing at the center of the flashing target. A 1-second rest period was provided for online feedback, marked by a red rectangle covering the target. Data were similarly collected using a Synamps2 EEG system, with a sampling rate of 1000 Hz and downsampling to 250 Hz. The same 50 Hz notch filter was employed for data processing.

\subsection{The Baseline Methods}
\label{sec:baseline}
Five baseline methods, for which code is available, were selected for comparison in this study. These methods include enhanced Canonical Correlation Analysis (eCCA) \cite{nakanishi2014high, chenHighspeedSpellingNoninvasive2015a}, Task-Related Component Analysis (TRCA) \cite{nakanishi2017enhancing}, Standard Canonical Correlation Analysis with QR decomposition (SCCA\_qr) \cite{gander1980algorithms}, EEGNet \cite{lawhernEEGNetCompactConvolutional2018}, and Convolutional Complex Neural Networks (CCNN) \cite{ravi2020comparing}.

\subsubsection{TRCA}
TRCA has garnered significant attention in SSVEP research due to its robust performance in terms of accuracy and information transfer rate. This method extracts task-related components by maximizing reproducibility during the task period. Like eCCA, a template for each frequency is created by averaging the EEG data corresponding to that frequency. Spatial filters are then derived using the TRCA method. The correlation coefficients between the projected EEG and various reference signals are calculated to determine the result.

\subsubsection{eCCA}
eCCA enhances the Canonical Correlation Analysis by introducing subject-specific reference signals. In this study, reference signals were derived by averaging the EEG signals from a source subject, which were then used to train a group of spatial filters for CCA decoding. The spatial filters produce several correlation scores, which are combined using a weighted summation approach \cite{wong2021transferring} to determine the target frequency with the highest score.

\subsubsection{CCNN}
CCNN differs from traditional methods by using the complex spectrum of EEG data, calculated using Fast Fourier Transform (FFT), as input. The CCNN architecture includes two layers: a spatial filter kernel and a horizontal kernel for feature extraction. The spatial filter projects the multi-channel complex spectrum into a weighted spectrum vector. Deep features are subsequently extracted using the horizontal kernel, with the final output derived from a fully connected and softmax layer.

\subsubsection{EEGNet}
EEGNet is a versatile deep-learning framework tailored for various EEG classification tasks. It comprises three 1-D convolutional layers: a vertical spatial kernel to enhance channel-specific information, a depth-wise kernel acting as a bandpass filter, and a final layer for feature extraction. The classification results are obtained through a softmax layer.

\subsubsection{SCCA\_qr}
The SCCA\_qr method employs QR decomposition to transform each data matrix into orthogonal ($Q$) and triangular ($R$) matrices. This reduces dimensionality and simplifies the data structure. CCA is applied to these reduced-dimensional $Q$ matrices to identify pairs of canonical variables with maximal correlation.

All baseline methods rely on supervised learning. A leave-one-out cross-validation strategy was implemented for evaluation, utilizing the data of one subject as the validation set and the data from the remaining subjects as the training set.

\subsection{Implementation Details}

Our training framework is structured to operate on epoch-based iterations, with a designated cap of 800 epochs for the training duration. We utilize the AdamW optimizer \cite{loshchilov2018decoupled}, noted for its adept handling of sparse gradients and the incorporation of a weight decay coefficient set to 0.05, enhancing regularization efforts. This optimizer is configured with adaptive learning rates, specifically with beta coefficients of 0.9 and 0.95, which control the exponential decay rates of the moving averages. Moreover, we incorporate an epsilon value of $1 \times 10^{-8}$ to ensure numerical stability during computations. 

A key feature of our approach is the implementation of a cosine decay learning rate scheduler that initiates with a warm-up phase. This phase is crucial as it gradually increases the learning rate, transitioning smoothly into a cosine decay pattern. This scheduling strategy enables precise adjustments to the learning rate throughout the training process. 

Our training dataset employs a random window sliding technique to enhance the model’s adaptability, generating 12,000 random window samples per epoch. This method effectively mitigates phase variability issues, significantly improving model robustness and facilitating more effective transfer learning. The diverse training scenarios created through random window sampling are key in acquiring domain-specific knowledge, thereby equipping the model with the necessary adaptability for robust performance across various environments. Additionally, the test set is also generated using random window selection to better evaluate the model’s transfer capacity. 

The effective learning rate, following previous work\cite{chenImprovingMaskedAutoencoders2023a,heMaskedAutoencodersAre2022}, \( lr \), is calculated using the formula:
\(lr = \textit{base\_lr} \times \frac{\textit{batch\_size} \times 2}{256}, \)
which ensures that the learning rate scales proportionally with changes in batch size. This method optimizes the training dynamics to suit our computational resources and the dataset's needs, aligning with our study's experimental schema.

The number of rules is based on the experiments of Section V-C2, and the order of the filers is based on Section V-C1. We show the results from the best model hyperparameter set in all Tables.

\begin{figure}
    \centering
    \includegraphics[width=1\linewidth]{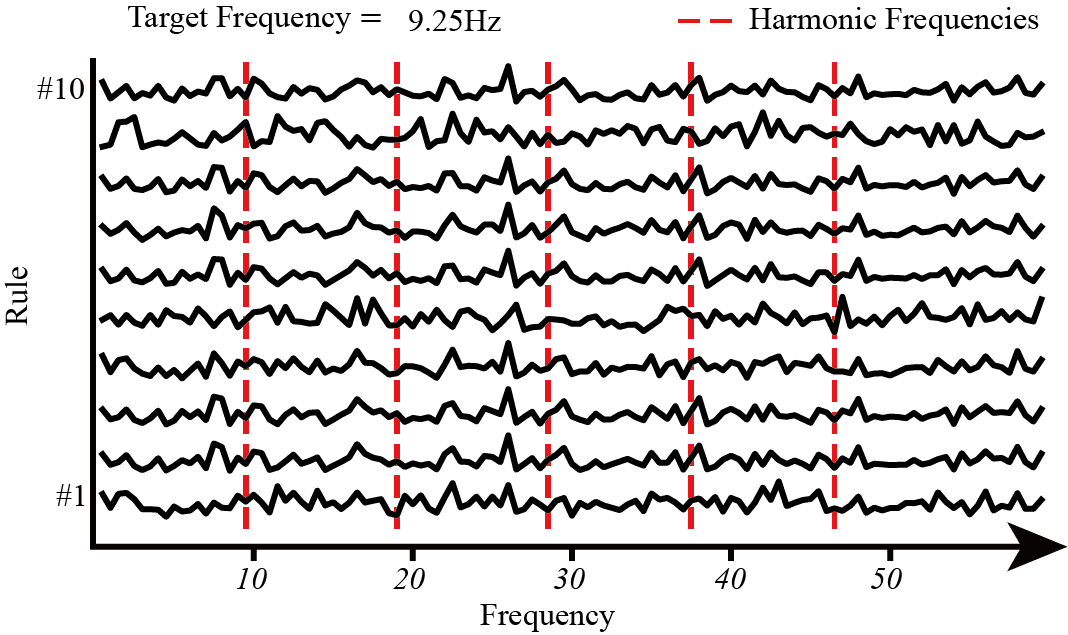}
    \caption{Incorrect case illustrating the failure identification of harmonic peaks by fuzzy rules.}
    \label{fig:inc_case}
\end{figure}

\begin{table*}[hp]
\centering
\captionsetup{width=\textwidth} 
\caption{Average ITR (bits/min) in bits per minute across subjects for six methods at varying data lengths on Dataset 12JFPM. Asterisks denote significant differences between iFuzzyTL and the other methods, determined by paired t-tests (*$p < 0.05$, **$p < 0.01$, ***$p < 0.001$).}
\label{tab:itr_12jfpm}
\fontsize{5}{6}\selectfont
\setlength{\tabcolsep}{3pt} 
\begin{tabular}{@{}cllllllllll@{}}
\toprule
Model & 0.5s & 0.6s & 0.7s & 0.8s & 0.9s & 1.0s & 1.1s & 1.2s & 1.5s & 2.0s \\ \midrule
TRCA & 8.30 (7.97) *** & 13.85 (13.12) *** & 24.66 (22.50) *** & 32.66 (28.75) *** & 45.80 (34.60) *** & 57.11 (42.66) *** & 61.54 (42.15) *** & 66.04 (43.31) *** & 75.56 (40.94) ** & 72.91 (29.85) * \\
eCCA & 94.42 (57.19) ** & 104.70 (64.91) ** & 104.01 (64.76) ** & 104.14 (65.34) ** & 108.63 (63.18) ** & 105.52 (58.41) *** & 103.11 (56.37) ** & 100.13 (50.43) ** & 93.43 (39.95) * & 76.75 (28.67) \\
CCNN & 144.46 (78.57) & 141.65 (75.38) & 137.61 (71.92) * & 139.40 (64.06) & 140.85 (56.86) & 133.47 (52.35) & 129.90 (45.17) & 125.29 (37.38) & 110.43 (25.62) & 90.44 (15.49) \\
EEGNET & 86.97 (53.45) ** & 97.61 (65.20) ** & 104.48 (57.76) *** & 105.11 (60.27) *** & 105.69 (56.63) *** & 109.04 (55.25) ** & 106.42 (50.57) ** & 104.51 (49.44) * & 99.45 (33.24) & 81.71 (23.05) \\
SCCA\_qr & 107.17 (73.31) & 116.88 (74.85) * & 122.10 (74.27) ** & 128.79 (76.01) * & 128.52 (70.73) * & 123.34 (64.28) * & 121.56 (59.84) & 117.71 (52.52) & 104.21 (35.71) & 87.80 (22.08) \\
iFuzzyTL & 148.30 (49.19) & 159.36 (57.45) & 168.06 (59.95) & 160.49 (51.90) & 160.28 (49.68) & 149.58 (43.73) & 138.66 (38.17) & 128.22 (35.22) & 110.74 (30.91) & 85.64 (22.85) \\
\bottomrule
\end{tabular}
\end{table*}

\begin{table*}[hp]
\centering
\captionsetup{width=\textwidth} 
\caption{Average ITR (bits/min) in bits per minute across subjects for six methods at varying data lengths on Dataset eldBETA. Asterisks denote significant differences between iFuzzyTL and the other methods, determined by paired t-tests (**$p < 0.01$, ***$p < 0.001$).}
\label{tab:itr_eldBETA}
\fontsize{5}{6}\selectfont
\setlength{\tabcolsep}{3pt} 
\begin{tabular}{@{}cllllllllll@{}}
\toprule
Model & 0.5s & 0.6s & 0.7s & 0.8s & 0.9s & 1.0s & 1.1s & 1.2s & 1.5s & 2.0s \\ \midrule
TRCA & 43.03 (44.56) *** & 42.67 (41.93) *** & 41.53 (39.99) *** & 42.33 (38.70) *** & 43.71 (37.15) *** & 42.73 (35.76) *** & 43.45 (34.59) *** & 42.71 (33.75) *** & 40.62 (29.56) *** & 33.30 (23.73) *** \\
eCCA & 55.15 (50.60) *** & 58.91 (48.40) *** & 62.84 (48.80) *** & 63.96 (47.76) *** & 66.20 (45.99) *** & 66.82 (44.57) *** & 66.55 (41.26) *** & 65.97 (38.27) *** & 61.76 (31.32) *** & 54.01 (23.55) *** \\
CCNN & 114.20 (77.99) & 107.73 (69.07) & 106.29 (64.25) *** & 101.58 (59.11) & 97.85 (55.03) & 93.58 (49.37) & 89.16 (45.66) ** & 85.37 (42.07) & 75.81 (33.22) ** & 62.20 (24.71) ** \\
EEGNET & 97.57 (71.88) *** & 91.85 (68.87) *** & 88.70 (64.41) *** & 84.86 (58.18) *** & 81.19 (54.79) *** & 77.11 (50.48) *** & 73.78 (47.08) *** & 69.87 (44.28) *** & 61.05 (36.14) *** & 49.77 (27.61) *** \\
SCCA\_qr & 42.57 (45.99) *** & 53.02 (47.24) *** & 57.28 (47.21) *** & 60.06 (45.67) *** & 61.81 (44.17) *** & 63.17 (42.29) *** & 63.23 (40.08) *** & 62.51 (38.24) *** & 59.73 (31.83) *** & 54.34 (23.72) *** \\
iFuzzyTL & 122.60 (64.10) & 107.90 (57.50) & 118.92 (57.71) & 99.52 (52.93) & 96.99 (49.29) & 94.63 (44.87) & 96.28 (42.29) & 86.54 (40.32) & 80.59 (31.23) & 65.76 (23.17) \\
\bottomrule
\end{tabular}
\end{table*}

\begin{table*}[hp]
\centering
\captionsetup{width=\textwidth} 
\caption{Average ITR (bits/min) in bits per minute across subjects for seven methods at varying data lengths on Dataset Benchmark. Asterisks denote significant differences between iFuzzyTL and the other methods, determined by paired t-tests (*$p < 0.05$, **$p < 0.01$, ***$p < 0.001$).}
\label{tab:itr_benchmark}
\fontsize{5}{6}\selectfont
\setlength{\tabcolsep}{3pt} 
\begin{tabular}{@{}cllllllllll@{}}
\toprule
Model & 0.5s & 0.6s & 0.7s & 0.8s & 0.9s & 1.0s & 1.1s & 1.2s & 1.5s & 2.0s \\ \midrule
TRCA & 78.79 (73.27) *** & 88.28 (78.56) *** & 97.41 (78.74) *** & 106.78 (78.09) *** & 111.31 (76.05) *** & 112.22 (73.57) *** & 109.10 (68.76) *** & 104.97 (65.05) *** & 96.84 (54.04) *** & 90.06 (40.82) *** \\
eCCA & 94.80 (66.85) *** & 117.86 (74.63) *** & 142.89 (73.63) *** & 154.86 (69.12) *** & 160.36 (62.77) *** & 159.76 (58.44) *** & 155.28 (55.03) *** & 154.18 (48.73) *** & 141.39 (35.66) *** & 123.05 (18.85) *** \\
CCNN & 176.82 (92.86) ** & 200.17 (92.49) & 206.75 (87.72) & 204.42 (79.06) & 195.95 (73.00) ** & 188.21 (66.14) ** & 182.57 (60.30) *** & 173.57 (55.98) *** & 156.21 (43.29) * & 127.67 (29.32) * \\
EEGNET & 117.85 (81.26) *** & 129.55 (83.98) *** & 147.60 (84.13) *** & 153.65 (81.28) *** & 154.24 (73.02) *** & 145.86 (69.27) *** & 139.62 (63.67) *** & 136.06 (59.92) *** & 121.10 (49.25) *** & 107.77 (36.67) *** \\
SCCA\_qr & 29.84 (28.56) *** & 49.32 (41.57) *** & 72.60 (52.28) *** & 94.36 (58.08) *** & 108.04 (58.48) *** & 117.44 (56.65) *** & 122.88 (53.74) *** & 127.44 (52.13) *** & 125.49 (42.37) *** & 114.26 (27.71) *** \\
iFuzzyTL & 219.41 (39.35) & 206.73 (51.44) & 204.42 (62.23) & 218.94 (62.34) & 222.14 (59.01) & 213.99 (58.62) & 209.39 (46.19) & 198.01 (39.66) & 169.97 (28.52) & 136.57 (12.88) \\
\bottomrule
\end{tabular}
\end{table*}

\bibliographystyle{IEEEtran}
\bibliography{ref, ref_Beining}